\documentclass[twocolumn,showpacs,preprintnumbers,amsmath,amssymb]{revtex4}
\usepackage{graphicx}
\usepackage{dcolumn}
\usepackage{bm}
\def\bea {\begin{eqnarray}}
\def\eea {\end{eqnarray}}

\def\be {\begin{equation}}
\def\ee {\end{equation}}

\begin{document}
\title{Negative binomial multiplicity distribution in proton-proton collisions in limited pseudorapidity intervals at LHC up to $\sqrt {s}$= 7 TeV and the clan model}
\author{Premomoy Ghosh}
\email{prem@vecc.gov.in}
\medskip
\affiliation{Variable Energy Cyclotron 
Centre, 1/AF Bidhan Nagar, Kolkata 700 064, India}            
\date{\today}

\begin{abstract}
Experiments at the Large Hadron Collider (LHC) have measured multiplicity distributions in proton-proton collisions at a new 
domain of center-of-mass energy ($\sqrt {s}$) in limited pseudorapidity intervals. We analyze multiplicity distribution data 
of proton-proton collisions at LHC energies as measured by the Compact Muon Solenoid (CMS) experiment in terms of characteristic 
parameters of the Negative Binomial Distribution (NBD) function that has played a significant role in describing multiplicity 
distribution data of particle production in high energy physics experiments, in the pre-LHC energy-range, in various kinds of collisions 
for a wide range of collision energy and for different kinematic ranges. Beside a single NBD, we apply the formalism of 
weighted superposition of two NBDs to examine if the multiplicity distribution data of CMS could be better explained. The weighted
superposition of two NBDs indeed explain the distribution data better at the highest available LHC energy and in large interval 
of phase space. The two-NBD formalism further reveals that the energy invariance of the multiplicity distribution of 
the ``soft'' component of particle production in hadronic collisions is valid at LHC also, as it is at RHIC and Tevatron.
We analyze the data further in terms of clan parameters in the framework of the two-NBD model. 

\end{abstract}

\maketitle
\section{Introduction}
Experimental study of multiparticle production in high-energy hadronic (proton-proton, $pp$ or proton-antiproton, $p\bar p$) 
collisions has reached a new high, in terms of energy, at the Large Hadron Collider (LHC) \cite{ref01}. A fast growth in 
energy of collisions could be possible due to significant advancement of collider technology in the last few decades. From 
tens of GeV \cite{ref02} at Intersecting Storage Ring (ISR), hundreds of GeV \cite{ref03} at Super Proton Synchrotron 
(SPS), both at CERN, finally the center-of-mass energy ($\sqrt {s}$) of collisions has reached thousands of GeV first at 
Tevatron at Fermilab \cite{ref04} and then at LHC at CERN \cite{ref05,ref06,ref07,ref08,ref09}. Remarkably, for this wide 
range of collision energy, the two-parameter Negative Binomial Distribution (NBD) function, as given below in Eq. - (1), 
played major role in describing multiplicity distributions of produced charged particles.

\begin{equation}
  P(n,\langle n \rangle, k) = \frac{\Gamma(k+n)}{\Gamma (k)\Gamma(n+1)}\left[\frac{\langle n \rangle}{k+\langle n \rangle}\right]^n \times \left[\frac{k}{k+\langle n \rangle}\right]^k
\end{equation}  

where $\langle n \rangle$ is the average multiplicity and the parameter $k$ is related to dispersion $D$, ($D^2 = \langle n^2 \rangle - \langle n \rangle^2$) 
by
\begin{equation}
\frac {D^2}{\langle n \rangle^2} = \frac{1}{\langle n \rangle} + \frac {1}{k}.
\end{equation}  
The charged particle multiplicity distributions in $pp$ collisions at the ISR energies and in $p\bar p$ collisions 
at $\sqrt {s}$ = 540 GeV at the SPS, fit with NBD function satisfactorily in the full pseudorapidity, $\eta$ (where 
$\eta = -ln[tan(\theta/2)]$ and  $\theta$ is the polar angle of the particle with respect to the counterclockwise 
beam direction) space as well as in limited pseudorapidity intervals. But, at $\sqrt {s}$= 900 GeV SPS energy, a single 
NBD function could describe the data only for small pseudorapidity intervals at the mid-rapidity region, while for larger 
intervals, where shoulder-like structure appeared in the multiplicity distribution, a single NBD function turned out to be inadequate. 
Appearance of sub-structures in multiplicity distributions at higher energies and in larger pseudorapidity intervals has been 
attributed \cite{ref10,ref11,ref12,ref13} to weighted superposition or convolution of more than one functions representing 
more than one source or process of particle productions. Such sub-structure in SPS data at $\sqrt {s} =$ 900 GeV and in Tevatron 
data at $\sqrt {s} =$ 1.8 TeV could be well explained by weighted superposition of two NBD functions \cite{ref11}. The NBD is quite
pertinent for $pp$ collisions at energies available at the Large Hadron Collider (LHC) also, as has been reported \cite {ref05,ref06} 
first by 'A Large Ion Collider Experiment' (ALICE). 

The study of high-energy particle collisions in pre-LHC energy-range, where soft processes of particle productions dominate, 
barring application of perturbative quantum chromodynamics (pQCD), depends mostly on phenomenological models. Many of these 
models \cite{ref10,ref11,ref12,ref13,ref14,ref15,ref16,ref17,ref18,ref19,ref20,ref21,ref22}, which deal with multiplicity 
distributions of produced particles, interpret matching of various data of multiplicity distributions involving NBD function 
well within respective framework. Sometimes, the $k$-parameter in NBD has very different meanings in some of these so far 
successful approaches indicating that the very wide occurrences of NBD in high energy experiments is not yet a well understood 
phenomenon. In the given scenario, a detailed study of multiplicity distribution data of $pp$ collisions at new LHC energies in 
terms of NBD would be worth carrying out for better understanding of the role of NBD in multiparticle production. 

\section{Multiplicity Distributions in Proton-proton collisions at LHC}
\label{}
At LHC, multiplicity distributions in proton-proton collisions at center-of-mass energies $\sqrt {s}$ = 0.9, 2.36 and 7 
TeV have been measured by different experiments, in different kinematic ranges and for different classes of events. All 
these LHC-experiments find that the mean multiplicities at the new LHC energies ($\sqrt {s}$= 2.36 and 7 TeV) had been 
underestimated by the event generators (like PYTHIA, PHOJET etc.) in use.

The ALICE has measured primary charged particles at $\sqrt {s}$ = 0.9 and 2.36 TeV in the mid-$\eta$ region in three 
limited overlapping $\eta$-intervals $|\eta|\leq \eta_c$ = 0.5, 1.0 and 1.3 \cite{ref05}, in non-single diffractive (NSD) 
inelastic proton-proton collisions. At $\sqrt {s}$ = 7 TeV, instead of NSD inelastic events, ALICE analyzed \cite{ref06}  
an event class requiring at least one charged particle in $|\eta |<1$ and measured multiplicity distribution in that 
$\eta$-interval only. The distributions measured by ALICE in the three $\eta $-intervals, $|\eta|\leq \eta_c$ = 0.5, 1.0 
and 1.3, at the two energies, $\sqrt{s} = 0.9$ TeV, $\sqrt{s} = 2.36$ TeV have been reported \cite{ref05} to match fairly 
well with NBD. The NBD fit to the distribution at $\sqrt {s}$= 7 TeV, measured by ALICE, has been reported \cite{ref06} to 
be slightly underestimating the data at low multiplicity ($n <5$) and slightly overestimating the data at high multiplicity 
($n >55$). The Compact Muon Solenoid (CMS) experiment has measured primary charged hadrons for all the three LHC energies 
in non-single diffractive (NSD) inelastic proton-proton collisions in the mid-$\eta_{cm}$ region in five overlapping 
$\eta$-intervals $|\eta|\leq \eta_c$ = 0.5, 1.0, 1.5, 2.0 and 2.4 \cite{ref07} around the center-of-mass pseudorapidity 
($\eta_{cm}=0$). The CMS experiment did not fit the distributions with NBD or other distribution functions but reported 
a change of slope in $P_{n}$ for $n>20$ in its largest $\eta$-interval, $\eta_{c}<2.4$. This feature becomes more pronounced 
with increasing $\sqrt{s}$. A Toroidal LHC Apparatus (ATLAS) experiment has measured \cite{ref08} charged particle multiplicities 
for different event classes characterized by different lower cuts on the number of charged particles ($n_{ch} < 1$, 2 and 6) in 
different kinematic ranges ($p_{T}>$ 100 MeV, 500 MeV in $|\eta|<$2.5). In contrast to ALICE, CMS and ATLAS experiments, which 
measured multiplicity distributions in the central region in pseudorapidity, the Large Hadron Collider Beauty (LHCb) experiment 
at LHC has its detector with geometrical acceptance limited in the forward region. The LHCb experiment has analyzed multiplicity 
distributions for hard interaction events (at least one long track with transverse momentum, $p_T >$ 1 GeV/c) from $pp$ interactions 
at $\sqrt {s}$= 7 TeV in non-overlapping pseudorapidity bins of width $|\eta|<$ 0.5 in the pseudorapidity range 2.5 $< \eta <$ 4.5.
 
It is important to note, at this point, that a few of the phenomenological models have already been contrasted with the LHC data. To 
explain the appearance of sub-structure in the distribution, the set of CMS data of charged hadron multiplicity has been 
analyzed \cite{ref21} in the framework of Independent Pair Parton Interaction (IPPI) \cite{ref22} which shows that the number of 
soft pair parton interactions from colliding particles and so the density of the partonic medium is large for the LHC data and 
increases with energy. Within the framework of the IPPI model, the findings favor enlarged role of collective effects in $pp$ 
collisions at LHC. Similar conclusion is obtained \cite{ref21} from analysis of the data in terms of Quark Gluon String Model 
(QGSM) \cite{ref23,ref24} that fits better to the data than the IPPI model. Another attempt \cite{ref25} to describe the multiplicity 
distribution data at the new LHC energies has been in terms of a two-component model in quantum statistical approach \cite{ref10} which 
described hadronic collision data up to $\sqrt {s}$ = 540 GeV. The model considers convolution of a NBD (with $k$=1) function and a 
Poisson Distribution (PD)function representing two components of source of particle productions: a thermally equilibrated chaotic 
source and another coherent source, respectively. The study revealed poor agreement between the data and the model. However, normalized 
moments of multiplicity distributions for the LHC data have been reproduced \cite{ref26} by a model considering measured distribution 
as superposition of a PD describing particle emission from one source and a NBD describing distribution of other sources.

\section{Objective}
\label{}
Our objective is a detailed study of multiplicity distributions of proton-proton collisions at available LHC energies in limited 
pseudorapidity intervals in terms of behavior of the characteristic parameters of NBD with respect to changing width of pseudorapidity 
interval, $|\eta |$ and center-of-mass energy, $\sqrt {s}$ of collision. 

Of the LHC data, we chose to analyze the data recorded by the CMS experiment \cite{ref07}, for the reason that, (a) contrary to the other 
LHC experiments, CMS experiment has measured multiplicity distributions for all the three LHC energies ($\sqrt {s}= 0.9$, 2.36 and 7 TeV), 
available so far, in similar pseudorapidity intervals $\eta_{c}= 0.5$ to $\eta_{c} = 2.4$ (five intervals) and for the same class 
(non-single-diffractive or NSD) of events, facilitating systematic study of dependence on $\eta_{c}$ and $\sqrt {s}$ for same class
of events with more data points. (b) detailed study of CMS data in terms of NBD function is still absent. (c) the phenomenological studies 
\cite {ref21, ref25, ref26} with published data of multiplicity distributions from $proton-proton$ collisions at LHC energies have primarily 
dealt with the CMS data and analysis of same data by different phenomenological models help comparison of models. 

We analyze the multiplicity distribution data first with single NBD function and then extend our study with weighted superposition of 
two NBD functions \cite{ref11}, where single NBD resulted in poor agreement with data. Analyzing the data in terms of weighted 
superposition of two distribution functions becomes pertinent because of the reported \cite{ref07} change in slope or appearance of 
sub-structure by the CMS experiment in multiplicity distribution for $n>20$ in its largest $\eta$-interval, $\eta_{c}<2.4$. According 
to the two-component model of Ref.- \cite{ref11}, the shoulder-like structure in the multiplicity distribution of hadronic collisions 
at $\sqrt {s}$= 0.9 and 1.8 TeV could be explained by weighted superposition of two NBDs, representing two classes of events, 
``semihard - events with minijets or jets'' and ``soft - events without minijets or jets''. 

Analysis in terms of weighted superposition of two NBDs could be carried out following the formalism \cite{ref11,ref27} of 
Giovannini and Ugoccioni for the LHC data, as has been suggested \cite{ref27} by the authors. Beside proposing \cite{ref11} the model to 
describe multiplicity distribution with sub-structure, Giovannini and Ugoccioni, predicted possible scenarios for structure of the ``soft'' 
and the ``semihard'' components of particle production in the full phase space in hadronic collisions in the TeV energy-range, in terms of the 
$k$-parameters of the two-NBD and the clan parameters \cite{ref28}. The study of multiplicity distribution and the clan structure 
analysis \cite{ref28} in terms of the two-component model was then extended \cite{ref27} to the limited pseudorapidity intervals 
for the situation when analysis in limited phase space is carried out after classification of events into ``soft'' and ``semihard'', 
to ensure that the weight factors for the components in the limited phase space remain same as those in the full phase. 
The resulting function of the weighted superposition of two NBDs is given by:
\begin{eqnarray}
&&P_{n}(\sqrt {s},\eta_{c}) = \alpha_{soft}(\sqrt {s}) \nonumber\\
&&P_{n}[\langle n \rangle_{soft}(\sqrt {s},\eta_{c}),{k}_{soft}(\sqrt {s},\eta_{c})] + [1 - \alpha_{soft}(\sqrt {s})] \nonumber\\ 
&&P_{n}[\langle n \rangle_{semihard}(\sqrt {s},\eta_{c}),{k}_{semihard}(\sqrt {s},\eta_{c})]
\end{eqnarray}
where $\alpha_{soft}$ is the fraction of ``soft'' events and is a function of $\sqrt {s}$ only. The other parameters, functions of both, 
the $\sqrt {s}$ and the $\eta_{c}$, have usual meanings as described for Eq. - (1) with suffixes in parameters indicating respective 
components.

In the discussed model, the parameter ${k}_{soft}$ is constant with energy of hadronic collisions, indicating 
validity of KNO-scaling \cite{ref31} for the ``soft'' component. In terms of  ${k}_{semihard}$, for the ``semihard'' component, 
which is expected to be dominant in the TeV energy range of LHC, the model proposes three scenarios: 1) the ``semihard'' component 
also follows the KNO-scaling, where ${k}_{semihard}$ remains constant with energy 2) ${k}_{semihard}$ decreases linearly with 
increasing energy indicating violation of KNO-scaling by the ``semihard'' component and 3) a QCD-inspired scenario, where the KNO-violation 
is not as strong as in scenario 2, ${k}_{semihard}$ starts decreasing with energy but asymptotically tends to a constant value. The 
energy dependence of $\langle n \rangle_{soft}$ and $\langle n \rangle_{semihard}$ in the TeV energy range has been extrapolated 
\cite{ref11} empirically from the data in the GeV energy domain. 

By fitting the LHC data with two-NBD function, constrained with the predictions for $\langle n \rangle_{soft}$ and $k_{soft}$  as in Ref.\cite{ref27}, 
the parameters related to the ``semihard'' component could be obtained to match with the respective predictions. But, as the data, available 
with us for the analysis, contains both the ``soft'' and the ``semihard'' components and there is no published LHC-data, as yet, to analyze the ``soft'' and
``semihard'' sub-samples separately, the parametrization of Giovannini and Ugoccioni cannot be used. 

In the context of the superposition of the ``soft'' and the ``semihard'' components, some significant observations \cite {ref29,ref30,ref12} 
in collider experiments at SPS, Tevatron and RHIC are worth mentioning. The analysis \cite{ref29} of data of $p\bar p$ collisions at $\sqrt {s}$= 
630 and 1800 GeV by CDF experiment at Tevatron, Fermilab in two isolated sub-samples of ``soft'' and 
``hard'' events revealed invariance of properties of ``soft'' sub-sample as a function of $\sqrt {s}$. The energy invariance of 
dynamical mechanism of inelastic multiparticle production in ``soft'' $pp$ collisions has been observed \cite{ref30} to be valid at 
$\sqrt {s}$= 200 GeV by the STAR experiment at RHIC, BNL also. A comparative study \cite{ref12} of charged particle multiplicities 
arising from non-single diffractive inelastic hadronic collisions at $\sqrt {s}$= 30 GeV to 1800 GeV, including the data at collider 
energies at $\sqrt {s}$= 200 GeV to 1800 GeV in UA5 (SPS) and E735 (Tevatron) experiments, revealed that the multiplicity distribution 
data of collider energies deviate from much discussed KNO-scaling \cite{ref31}, which satisfactorily explains multiplicity distributions 
up to the ISR energies. The deviation, in the form of a shoulder-like structure apparently appeared in the collider data due to superposition 
of distribution of particles from some other process, different from the KNO producing process, on the top of the KNO distribution.

The ``hard'' events in the referred experimental analysis and the ``semihard'' events as termed in the discussed model are 
essentially similar class of events, involving hard parton-parton scatterings (due to high momentum transfer) resulting in QCD jets of
high transverse momentum above a certain threshold. 

In terms of two-NBD, we aim to study the energy dependence of the two components of particle production. Our interest lies particularly 
with the ``soft'' component of particle production in $pp$ collisions at LHC, in view of the energy invariance of the ``soft'' component 
observed in collider energies prior to LHC. We extend the study further to the clan structure analysis \cite{ref27,ref28} for the two 
components.

\section{Analysis and Discussions}
\label{}
\subsection{Behavior of the NBD parameters}
\label{}
\begin{table}[h]
\begin{center}
\label{tab}
\begin{tabular}{|c|c|c|c|c|}
\hline
$\sqrt{s}$ (TeV)& $\eta_{c}$& $\langle n \rangle_{NBD}$& $\langle n \rangle$& $\chi^2/d.o.f$ \\
\hline
$ 0.9$ & $0.5$ & $ 3.66\pm 0.04$ & $3.59^{+0.15}_{-0.15}$ & $  6.12/22$ \\
$ $   & $1.0$ & $ 7.49\pm 0.08$ & $7.26^{+0.16}_{-0.15}$ & $ 53.35/38$ \\
$ $   & $1.5$ & $11.32\pm 0.10$ & $10.95^{+0.18}_{-0.16}$ & $ 49.55/50$ \\
$ $   & $2.0$ & $15.26\pm 0.13$ & $14.83^{+0.21}_{-0.18}$ & $ 36.69/60$ \\
$ $   & $2.4$ & $18.36\pm 0.14$ & $17.86^{+0.23}_{-0.20}$ & $ 46.29/66$ \\
\hline
$2.36$ & $0.5$ & $ 4.70\pm 0.08$ & $4.60^{+0.16}_{-0.15}$ & $  6.38/21$ \\
$ $   & $1.0$ & $ 9.42\pm 0.11$ & $9.26^{+0.19}_{-0.17}$ & $ 55.30/38$ \\
$ $   & $1.5$ & $14.35\pm 0.16$ & $14.01^{+0.28}_{-0.21}$ & $ 24.79/48$ \\
$ $   & $2.0$ & $19.35\pm 0.21$ & $18.93^{+0.29}_{-0.27}$ & $ 29.11/58$ \\
$ $   & $2.4$ & $23.35\pm 0.25$ & $22.63^{+0.35}_{-0.33}$ & $ 29.76/68$ \\
\hline
$7$ & $0.5$ & $ 6.16\pm 0.05$ & $5.98^{+0.14}_{-0.13}$ & $ 83.36/39$ \\
$ $   & $1.0$ & $12.49\pm 0.08$ & $12.18^{+0.15}_{-0.13}$ & $152.65/68$ \\
$ $   & $1.5$ & $18.89\pm 0.10$ & $18.53^{+0.18}_{-0.15}$ & $226.57/93$ \\
$ $   & $2.0$ & $25.47\pm 0.14$ & $25.10^{+0.21}_{-0.19}$ & $208.56/113$ \\
$ $   & $2.4$ & $30.90\pm 0.16$ & $30.32^{+0.24}_{-0.21}$ & $129.37/125$ \\
\hline
\end{tabular}
\caption{Table showing $\langle n \rangle_{NBD}$ parameter of NBD, $\langle n \rangle$ calculated from published
multiplicity distribution and the $\chi^2 /d.o.f.$ for different pseudorapidity intervals for $\sqrt {s}$ = 0.9, 2.36 and 7 TeV.}
\end{center}
\end{table}
We fit the multiplicity distribution data \cite{ref07,ref32} for pseudorapidity intervals $\eta_{c}=$ 0.5, 1.0, 1.5, 2.0 and 2.4 with a single NBD function 
and tabulate mean multiplicity $\langle n \rangle$ calculated from the distribution data, the values of best fitted parameter, 
$\langle n \rangle_{NBD}$, and the corresponding values of $\chi^2/d.o.f$ in Table-1. 
The increase in average multiplicity, $\langle n \rangle$ with increasing energy and with the 
width of the symmetric pseudorapidity intervals around $\eta_{cm}=0$ in the mid-rapidity region
is a fact well established by experiments. 
As the parameter $\langle n \rangle$ of NBD function, 
given by Eq.- (1), gives the average multiplicity, the closeness of its best fitted values with respective 
measured mean multiplicity and its behavioral pattern with respect to $\sqrt {s}$ and $|\eta|$-intervals, as shown 
in Table-1, is expected. The behavior of $\langle n \rangle$-parameter of NBD with respect to $\sqrt {s}$ and 
$\eta_{c}$ for the analyzed data is more clearly depicted in Fig.1.    

The Fig.2 shows the dependence of the NBD-parameter, $k$ on the size of the pseudorapidity interval, $\eta_{c}$ and 
on the energy, $\sqrt {s}$ of collisions at LHC. In the considered energy domain, for a given rapidity interval, 
$k$ decreases with increasing center-of-mass energy of collision. This behavior is consistent 
with the observed energy dependence of $k$ in the pre-LHC energy range. 
\begin{center}
  \begin{figure}[h]
    \includegraphics[scale=0.38]{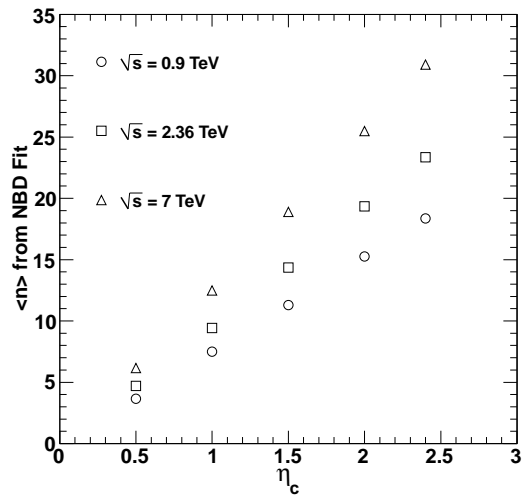} 
    \caption{Parameter $\langle n \rangle$ from NBD for $|\eta|<$0.5 to 2.4 for $\sqrt {s}$ = 0.9, 2.36 and 7 TeV. 
The error-bars associated with the data-points are not visible as the corresponding magnitudes are smaller than the dimension of symbol-size in the plots.} 
    \label{fig:etaVSn} 
  \end{figure}
\end{center}
\begin{center}
  \begin{figure}[h]
    \includegraphics[scale=0.38]{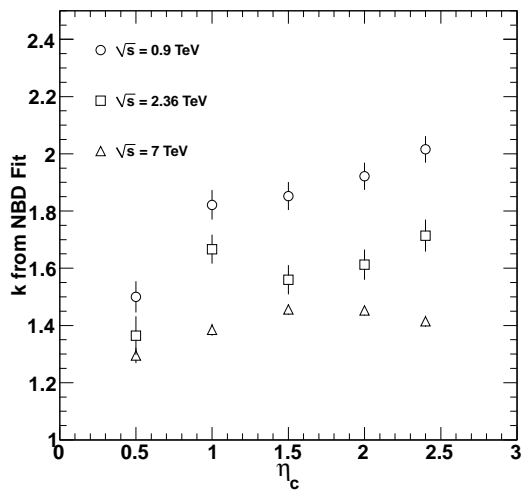} 
    \caption{Parameter $k$ of NBD for $|\eta|<$0.5 to 2.4 for $\sqrt {s}$ = 0.9, 2.36 and 7 TeV.} 
    \label{fig:parameters} 
  \end{figure}
\end{center}
On the other hand, for a given $\sqrt {s}$, 
though the general trend in the behavior of $k$ shows increase in $k$ with increase in the size of the symmetric 
pseudorapidity window around the center-of-mass pseudorapidity $\eta_{cm} = 0$, deviations appear at $\eta_{c}=1.0$ 
for the $\sqrt {s}=$ 0.9 and 2.36 TeV data as have been shown in the Fig.2. The reasons for such deviations 
is not understood. Further, the rate of increase in $k$ with increased size of the pseudorapidity interval decreases 
with increasing energy. 
\begin{center}
\begin{figure}[h]
\includegraphics[scale=0.38]{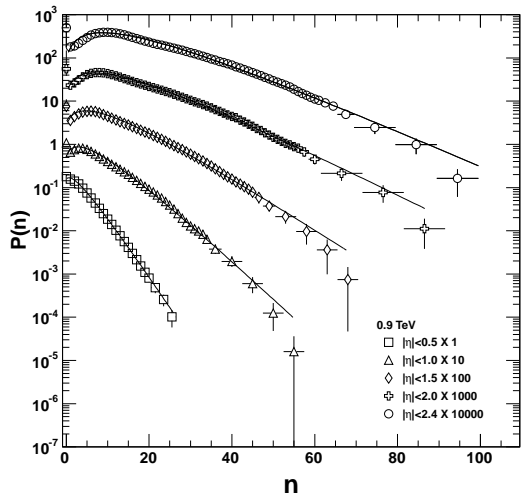} 
\caption{Primary charged hadron multiplicity distributions for $|\eta|<$0.5 to 2.4 for $\sqrt {s}$ = 0.9 TeV. 
The solid lines drawn along the data-points correspond to respective fits of NBD. The error-bars include both the statistical and the systematic uncertainties}
\label{fig:distribution1} 
\end{figure}
\end{center}
\begin{center}
\begin{figure}[h]
\includegraphics[scale=0.38]{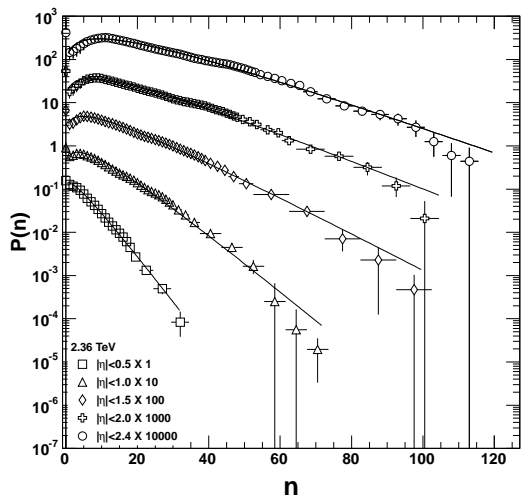} 
\caption{Primary charged hadron multiplicity distributions for $|\eta|<$0.5 to 2.4 for $\sqrt {s}$ = 2.36 TeV. 
The solid lines drawn along the data-points correspond to respective fits of NBD. The error-bars include both the 
statistical and the systematic uncertainties}
\label{fig:distribution2} 
\end{figure}
\end{center}
For $\sqrt {s} = $ 7 TeV, however, the trend is followed up to $|\eta|<$1.5 beyond which the trend gets reversed. 
The reason for this deviation may be due to the appearance of sub-structure in the multiplicity distributions 
and so the inadequacy of a single NBD function to fit the distributions as has already been reflected in terms of 
the $\chi^2/d.o.f$ values, as listed in Table - 1, for the fits to the distributions at $\sqrt {s}$ = 7 TeV. 
\begin{center}
\begin{figure}[h]
\includegraphics[scale=0.38]{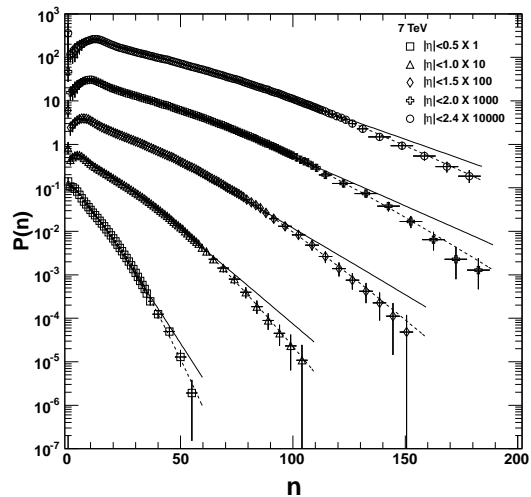} 
\caption{Primary charged hadron multiplicity distributions for $|\eta|<$0.5 to 2.4 for $\sqrt {s}$ = 7 TeV. 
The solid lines drawn along the data-points correspond to respective fits of single NBD while the dashed lines 
correspond to respective fits of Two-NBD. The error-bars include both the statistical and the systematic uncertainties}
\label{fig:distribution3} 
\end{figure}
\end{center}
We, therefore, fit the $\sqrt {s} = $ 7 TeV data with weighted superposition of two NBD functions (given by Eq. - (3)) also. 
We plot multiplicity distribution data for $\sqrt {s}= 0.9$ TeV (Fig.3), for $\sqrt {s}= 2.36$ TeV (Fig.4) along with best fitted 
NBD function and for $\sqrt {s}= 7$ TeV (Fig.5) along with best fitted NBD function and superposition of two NBD functions.

As could be seen in Fig.- 5, the multiplicity distribution data of charged hadrons for $\sqrt {s}$= 7 TeV for all the 
pseudorapidity intervals fit better to the weighted superposition of two-NBDs than to a single NBD. The improvement
is more clear in terms of $\chi^2/d.o.f$ values which, along with values of best fitted free parameters of the two-NBD 
function given by Eq. - (3), are tabulated in Table - II. 
\begin{table}[h]
\begin{center}
\label{tab}
\begin{tabular}{|c|c|c|c|c|}
\hline
${k}_{s}$ &$\langle n \rangle_{s}$& ${k}_{sh} $&$\langle n \rangle_{sh}$& $\chi^2/d.o.f$ \\
\hline
$1.30\pm0.29$ & $ 4.53\pm1.11$ & $5.61\pm4.17$ & $14.08\pm2.32$& $ 4.14/35$\\
$1.48\pm0.28$ & $ 8.90\pm1.67$ & $5.22\pm1.17$ & $26.37\pm3.26$& $ 6.97/64$\\
$1.73\pm0.28$ & $12.16\pm1.89$ & $4.74\pm0.87$ & $36.35\pm3.89$& $12.92/89$\\
$1.98\pm0.27$ & $14.86\pm1.89$ & $4.23\pm0.72$ & $44.87\pm4.12$& $14.93/109$\\
$2.38\pm0.34$ & $15.06\pm1.48$ & $3.25\pm0.49$ & $46.86\pm3.45$& $11.91/121$\\
\hline
\end{tabular}
\caption{Table showing best fitted parameters of function given by Eq. - (3) and corresponding $\chi^2 /d.o.f.$ for
multiplicity distributions in pseudorapidity intervals $\eta_{c}<$0.5, 1.0, 1.5, 2.0 and 2.4 (tabulated in order) for 
$\sqrt {s}$ = 7000 GeV. Suffixes $s$ and $sh$ to the title of columns represent $soft$ and $semihard$, respectively. 
Soft events fraction decreases from 84\% ($\eta_{c}<$0.5) to 49\% ($\eta_{c}<$2.4).}
\end{center}
\end{table}
From Table-II, we also observe systematic trend in $\eta_{c}$-dependence of the best fitted values of parameters of two-NBD, 
for $\sqrt {s}$ = 7 TeV data. We find $\langle n \rangle_{semihard} \approx$ 3$\langle n \rangle_{soft}$ for a given 
$\eta$-interval. ${k}_{soft}$ increases and ${k}_{semihard}$ decreases with increase in $\eta_{c}$. It may be noted 
that the values of $\chi^2/d.o.f$ become lower than unity which may be due to the fitting of the function considering 
data-points with large errors (include both statistical and systematic errors).

\subsection{Energy invariant multiplicity distribution of soft component}

As already discussed in Section - III, the study of multiplicity distributions in terms of two-NBD could be important in 
the context of the experimental observations \cite {ref29,ref30} of energy invariance of multiplicity distributions of the 
soft component of events. We have also discussed why the predictions \cite{ref27} by Giovannini and Ugoccioni on two-NBD parameters for 
the TeV energy-domain in limited phase space could not be used for our analysis. Alternately, the two-NBD with parameters related to 
the ``soft'' component constrained with low-energy data could have been used. But, as the two-NBD parameters are functions of both 
the $\sqrt {s}$ and the $\eta_{c}$, the study of dependence on $\sqrt {s}$ could be possible only in the same $\eta_{c}$ and vice versa.

So, to study the $\sqrt {s}$ and $\eta_{c}$ dependencies of the two components, we prefer to study the distribution data of CMS at 
$\sqrt {s}$ = 0.9, 2.36 and 7 TeV in $\eta_{c}<$ 0.5, 1.0, 1.5, 2.0 and 2.4 in terms of the two-NBD function given by Eq. - (3) with 
unconstrained parameters (keeping ${k}_{soft}$, $\langle n \rangle_{soft}$, ${k}_{semihard}$ and $\langle n \rangle_{semihard}$ all free 
to produce best fit). All these fittings resulted in lowering of $\chi^2/d.o.f$ (to different extents) as compared to fits with single NBD, 
although the improvement in the fit by two-NBD is not that significant in small $\eta_{c}$ at $\sqrt {s}$ = 0.9 and 2.36 TeV. It is 
important to note that fitting the distributions as measured by ALICE \cite{ref05} up to $\eta_{c}<$ 1.3 at $\sqrt {s}$ = 0.9 
and 2.36 TeV with sum of two NBDs also did not significantly improve the description of data. Also, the values of best fitted free parameters 
obtained from our two-NBD fits at small pseudorapidity intervals ($\eta_{c}$), do not follow any systematic trend with respect to $\sqrt {s}$ 
or $\eta_{c}$, in general, for the data at $\sqrt {s}$ = 0.9 and 2.36 TeV .

Interestingly, however, for the largest available $\eta$-interval, where shoulder-like structures appear \cite{ref07} in multiplicity distributions, 
the unconstrained free parameters of the two-NBD show systematic behavior with respect to $\sqrt {s}$, as can be seen in Table - III.
\begin{table}[h]
\begin{center}
\label{tab}
\begin{tabular}{|c|c|c|c|c|}
\hline
$\sqrt {s} (TeV)$&${k}_{s}$ &$\langle n \rangle_{s}$& ${k}_{sh} $&$\langle n \rangle_{sh}$\\
\hline
$0.9$ & $2.44\pm0.32$ & $14.78\pm1.99$ & $8.13\pm2.34$ & $35.11\pm3.90$\\
$2.36$ & $2.57\pm0.52$ & $15.74\pm2.98$ & $6.27\pm2.21$ & $41.92\pm6.21$\\
$7.0$ & $2.38\pm0.34$ & $15.06\pm1.48$ & $3.25\pm0.49$ & $46.86\pm3.45$\\
\hline
\end{tabular}
\caption{Table showing best fitted free parameters of function given by Eq. - (3) for
multiplicity distributions in pseudorapidity interval, $\eta_{c}<$2.4 for $\sqrt {s}$ = 0.9, 2.36 and 7 TeV. Suffixes $s$ and $sh$ to 
the title of columns represent $soft$ and $semihard$, respectively. Soft events fraction is 82\%  for $\sqrt {s}$ = 0.9 TeV, 69\%
for $\sqrt {s}$ = 2.36 TeV and 49\% $\sqrt {s}$ = 7 TeV.}
\end{center}
\end{table}
The lack of systematic trend in two-NBD parameters at small $\eta_{c}$ in contrast to that at large $\eta_{c}$, where the energy invariance 
of parameters related to soft component is observed in the LHC data, needs further discussion, particularly, as the energy-invariance of the ``soft'' component 
of particle productions has been observed in $\eta_{c}<$1.0 by both the STAR and the CDF experiments and the model involving superposition 
of two NBDs has been shown to be valid in small $\eta_{c}$-intervals. The different behavior of LHC data at small $\eta_{c}$-interval may be 
due to analysis of different class of event samples as compared to STAR and CDF. We recollect, the validity of the two component model 
\cite{ref11} of Giovannini and Ugoccioni for the full phase space was extended \cite{ref27} to limited pseudorapidity intervals only after 
classification of events is carried out in the full phase space, to ensure multiplicity distributions in limited intervals holding same
weighting factor as that in the full phase space. Also, both the STAR and the CDF experiments analyzed isolated ``soft'' and ``hard'' 
sub-samples of events separately. But, this work deals with available LHC data sample including both the ``soft'' 
and the ``semihard'' events for a given $\eta$-interval and attempts to extract contributions of the components by fitting multiplicity 
distribution data by weighted superposition of two NBDs. A small pseudorapidity interval may not include all the particles of an 
event and possibility of exclusion of part of an event in the small interval is more for high multiplicity events, which are likely to be 
abundant at new LHC-energies. So, the multiplicity distribution of a sample of all events, including the ``soft'' and the ``semihard'' components, 
in small interval of pseudorapidity may not reflect the same weighting factor for the two-NBD fit as that in the full phase space. The effect of 
partial exclusion of an event minimizes with increasing size of phase space. In a large pseudorapidity interval, therefore, the weighting
factor of the two-NBD may remain same as that in the full phase space. 

Nevertheless, application of the formalism of the weighted superposition of two NBDs in the analysis of the published multiplicity data of CMS 
reveals significant property of energy invariance of ``soft'' component of particle productions at LHC in the largest available
pseudorapidity interval, $\eta_{c}<$2.4, where shoulder-like structure appear in the multiplicity distributions. At this point, it is important to 
compare the goodness of the fits to the multiplicity distribution data by a single NBD and by the superposition of two-NBDs. To compare the goodness 
of fits at the considered LHC energies in the pseudorapidity interval, $\eta_{c}<$ 2.4, we carry out the residual analysis and plot the residuals in 
Fig. - 6. The residual is defined \cite{ref11} as the difference between a data-point and corresponding fit-value. It is clear from the plots 
in Fig. - 6, that the weighted superposition of two NBD functions fits better than a single NBD with the available multiplicity distribution 
data of LHC at $\sqrt {s}$ = 0.9, 2.36 and 7 TeV at $\eta_{c}<$ 2.4.  
\begin{center}
  \begin{figure}[h]
    \includegraphics[scale=0.40]{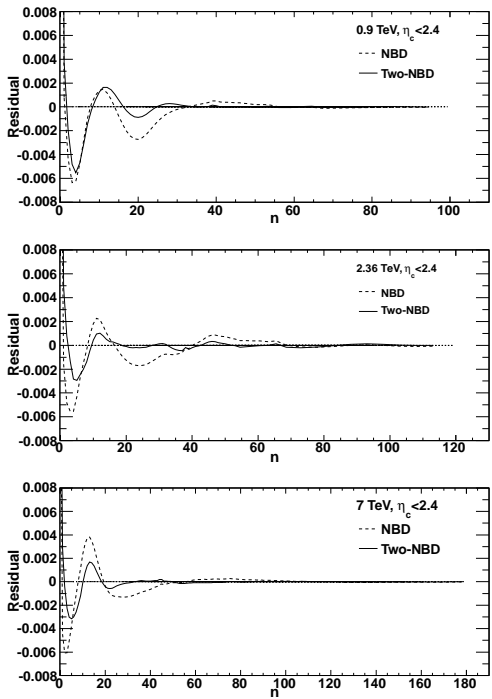} 
    \caption{Plots of residual analysis to test the goodness of fits for $\sqrt {s}$ = 0.9 TeV, $\sqrt {s}$ = 2.36 TeV and $\sqrt {s}$ = 7 TeV. The dotted lines correspond to fits to a single NBD function and the continuous lines correspond to fits to weighted superposition of two NBDs.}
    \label{fig:residual} 
  \end{figure}
\end{center}

\subsection{Clan structure analysis in the two-component model}

Better agreement of two-NBD with data at $\sqrt {s}$= 7 TeV in all available $\eta_{c}$ and at $\sqrt {s}$ = 0.9 and 2.36 TeV
in $\eta_{c}<$ 2.4, motivates us in contrasting the LHC data at TeV energies with the clan structure analysis within the framework 
of the two-component model \cite {ref11, ref27}. 

The discussed two-component model fits in the framework of clan structure analysis \cite{ref28}. In fact, the authors of Ref. \cite{ref27} 
have pointed out that to distinguish different scenarios, one should look at the $1/k$ parameters and related clan structure analysis. 
The clan model \cite {ref28} is based on cascading process, where particles are emitted from a previously produced particle while the producing 
particle can change its momentum and quantum numbers during the process, as it happens in case of well known fragmentation and decay processes. 
The group of particles including one originally produced from the collision, directly or indirectly, and particles produced from that in 
following steps of cascading form a $cluster$ with the originally produced first particle termed as the $ancestor$ of the cluster. Such a 
cluster or a group of particles with common ancestry is termed as $clan$ \cite {ref28}. As per definition, a clan contains at least one 
particle. Clans can be assumed to be produced independently. The characteristic parameters of the clan model are the average number of clans, 
$\bar N$ and the average number of charged particles per clan, $\bar n_{c}$, which are related to the NBD parameters as follows:     
\begin{equation}
\bar N = k\times ln(1 + \frac{\langle n \rangle}{k})
\end{equation}  
and 
\begin{equation}
\bar n_{c} = \frac{\langle n \rangle}{\bar N}
\end{equation}  
For the two-NBD two-component model, describing ``soft'' and ``semihard'' components of events, the behavior of the clan parameters 
need to be studied separately for each component.

For the study of $\sqrt {s}$-dependence of clan parameters, we consider the multiplicity distributions and corresponding fit-parameters 
of two-NBD in $\eta_{c}<$ 2.4 at $\sqrt {s}$ = 0.9, 2.36 and 7 TeV and plot the clan parameters, along with the respective NBD-parameter 
$1/k$, for both the ``soft'' and the ``semihard'' components in Fig. - 7. For the study of $\eta_{c}$-dependence of the same parameters, 
we consider multiplicity data in $\eta_{c}<$ 0.5, 1.0, 1.5, 2.0 and 2.4 at $\sqrt {s}$= 7 TeV and plot, in Fig. - 8, the parameters similar 
to those in Fig. - 7. The NBD-parameter $1/k$ is plotted along with the clan parameters for convenience of comparing the behavior of all the 
three parameters, $1/k$, $\bar n_{c}$ and the $\bar N$ with predictions of the two-component model \cite{ref11,ref27}.    

As can be seen from the plots in Fig. - 7, for the ``soft'' component, all the three parameters, characterizing different scenarios as 
particle productions according to the discussed model in Ref.-\cite{ref27}, the NBD parameter, $1/k$, the average number of charged particles 
per clan, $\bar n_{c}$ and the average number of clans, $\bar N$ are constant with energy. For the ``semihard'' component, $1/k$ and $\bar n_{c}$ 
increase while $\bar N$ decreases with increase in energy. The rate of change in the parameters is rapid in the range from 2.36 TeV to 7 TeV 
as compared to the that in the range from 0.9 TeV to 2.36 TeV. Comparing the nature of the energy-dependence of these parameters (though not for 
same $\eta_{c}$) as shown in Ref. \cite{ref27}, one sees that the behavior of energy dependence of clan parameters and the $k_{semihard}$-parameter 
of two-NBD in $\eta_{c<}$ 2.4 in the new LHC-energies match with the scenario - 2 (as discussed in Section - III) of the two-component model of 
particle production, represented by weighted superposition of two NBD, indicating violation of KNO scaling by the ``semihard'' component. 
But, contrary to the prediction by the model, $1/k_{soft}$ is always larger than $1/k_{semihard}$ in the considered energy-range.  
\begin{center}
  \begin{figure}[h]
    \includegraphics[scale=0.40]{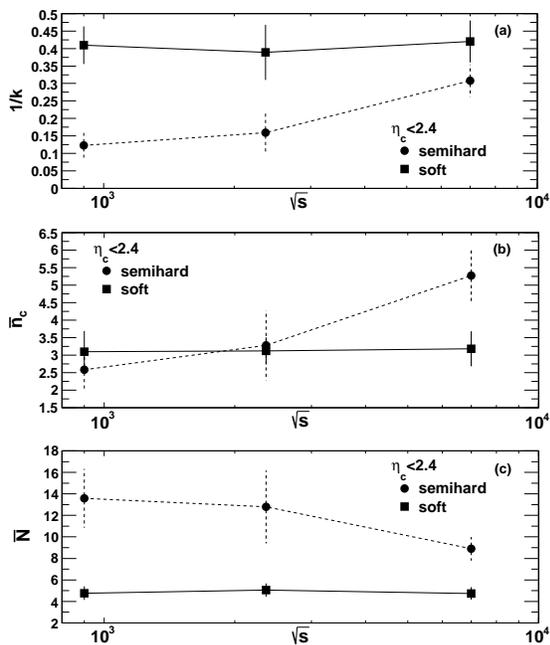} 
    \caption{Energy dependence of the NBD parameter $1/k$ (panel a), the average number of charged particles per clan, $\bar n_{c}$ (panel b) and 
the average number of clans, $\bar N$ (panel c) obtained for ``soft'' and ``semihard'' components from multiplicity distribution data in $|\eta|<$2.4 
in the LHC energy-domain ($\sqrt {s}$ = 0.9, 2.36 and 7 TeV). The lines in the plots are drawn joining the points to guide the eye.}
    \label{fig:clanlhcEnergy} 
  \end{figure}
\end{center}
The behavior of $\eta_{c}$-dependence, as depicted in Fig. - 8, of the characteristic parameters for the two components, however, do not corroborate
the finding from the study of the energy dependence. According to the model, the $1/k_{soft}$ decreases and $1/k_{semihard}$ decreases rapidly with 
increasing $\eta_{c}$ in the scenario - 2 of the model. The plot of $\eta_{c}$ - dependence of $1/k_{semihard}$ for the data at $\sqrt {s}$ shows rising
tendency.
\begin{center}
  \begin{figure}[h]
    \includegraphics[scale=0.40]{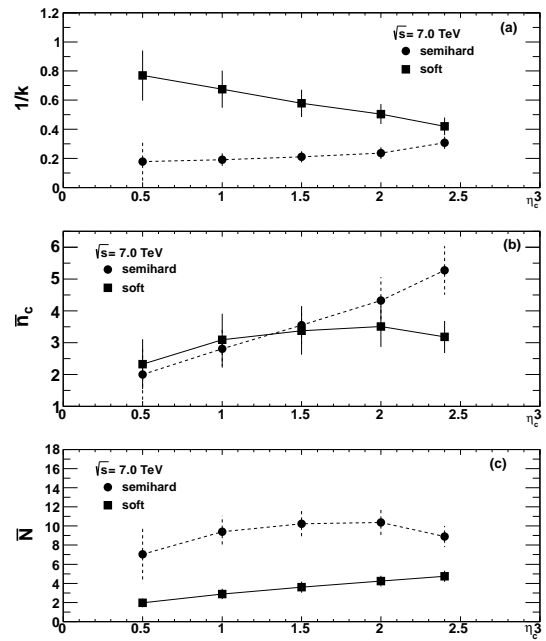} 
    \caption{The $\eta_{c}$-dependence of the NBD parameter $1/k$ (panel a), the average number of charged particles per clan, $\bar n_{c}$ (panel b) and 
the average number of clans, $\bar N$ (panel c) obtained for ``soft'' and ``semihard'' components from multiplicity distribution data at $\sqrt {s}$ = 
7 TeV. The lines in the plots are drawn joining the points to guide the eye.}
    \label{fig:clanlhcEta} 
  \end{figure}
\end{center}

\section{Summary}

Our study on multiplicity distributions of charged hadrons from $pp$ collisions in limited pseudorapidity intervals
at LHC energies in terms of Negative Binomial Distribution function reveals the followings:

1) At pseudorapidity intervals of small sizes, particularly at energies $\sqrt {s}$ = 0.9 and 2.36 TeV, a single 
NBD function fits the distribution data reasonably well, while parameters of two-NBD show no systematic trend. For 
the distribution data at $\sqrt {s}$ = 7 TeV, however, a single NBD function appears inadequate, while weighted 
superposition of two NBDs fit the data satisfactorily. 

2) The energy ($\sqrt {s}$) invariance of the parameters related to the ``soft'' component and so of the respective 
multiplicity distribution at $\eta_{c}<$ 2.4, where the measured distributions show shoulder-like structure. 
This observation could be indicative to the invariance of the dynamical mechanism of ``soft'' multiparticle production, 
as has already been seen by the STAR experiment in $pp$ collision data at $\sqrt {s}$ = 200 GeV and by the CDF experiment 
in $p\bar p$ collisions at $\sqrt {s}$ = 630 and 1800 GeV. 

3) The multiplicity distributions for all the available LHC energies in  $\eta_{c}<$ 2.4 agree better with weighted superposition of two 
NBDs than a single NBD function.  

4) Behavior of clan parameters show energy-dependence in accordance with one of the predicted scenarios by the two-source model. But the 
study of $\eta_{c}$-dependence of data at $\sqrt {s}$ = 7 TeV, which otherwise fit with two-NBD, does not substantiate the finding in the
energy-dependence study.

This study of multiplicity distributions of the $pp$ collisions at LHC, in terms of NBD, highlights significant features of mutiparticle productions 
in hadronic collisions at LHC energies. In spite of limitations in the available data in small pseudorapidity intervals vis-a-vis the adopted two-NBD 
formalism, the two-NBD describes the multiplicity distributions data better in all pseudorapidity intervals at $\sqrt {s}$= 7 TeV and in large 
pseudorapidity intervals at $\sqrt {s}$ = 0.9 and 2.36 TeV. The most striking revelation from the analysis, following the formalism of weighted 
superposition of two NBDs, is the energy invariance of multiplicity distribution of the ``soft'' component of events in the largest available 
pseudorapidity interval of LHC data, where the distributions show sub-structure. Importantly, the energy invariance has been observed by fitting
unconstrained two-NBD to the distribution data at different considered energies. In the context of the finding, the energy invariance, in this analysis 
and similar feature as observed in Fermilab and RHIC energies, in rapidity interval of smaller size, we suggest that the findings be corroborated with 
analysis of LHC data of isolated sub-samples of ``soft'' and ``hard (semihard)'' events, as has been studied by STAR at RHIC and CDF at Tevatron.
\section{Acknowledgements}
The author acknowledges useful discussions with Prithwish Tribedy and Sudipan De.

\end{document}